\documentclass[11pt,a4paper]{article}

\usepackage[utf8]{inputenc}
\usepackage[T1]{fontenc}
\usepackage{lmodern}
\usepackage{amsmath,amssymb,amsthm}
\usepackage{graphicx}
\usepackage{hyperref}
\usepackage{geometry}
\usepackage{enumitem}
\usepackage{booktabs}
\usepackage{array}   
\usepackage{tabularx}
\usepackage{multirow}
\usepackage{fancyhdr}
\title{Review of Blockchain-Based Approaches to Spent Fuel Management in Nuclear Power Plants}

\author{
  Yuxiang Xu\textsuperscript{1,2},\;
  Wenjuan Yu\textsuperscript{1,3},\;
  Yuqian Wan\textsuperscript{1},\;
  Zhongming Zhang\thanks{Corresponding author: zhongming.zhang@lancaster.ac.uk}\textsuperscript{1,2}
}

\date{%
  \textsuperscript{1}School of Engineering, Lancaster University, Lancaster, United Kingdom, LA1 4YW\\%
  \textsuperscript{2}Beijing Jiaotong University, Weihai Campus, Weihai, Shandong Province, China, 264401\\%
  \textsuperscript{3}School of Computing and Communications, InfoLab21, Lancaster University, United Kingdom, LA1 4YW\\[1ex]%
  \today
}

\begin{document}
\maketitle

% Abstract (Do not insert blank lines, i.e. \\) 
\abstract{This study addresses critical challenges in managing the transportation of spent nuclear fuel, including inadequate data transparency, stringent confidentiality requirements, and a lack of trust among collaborating parties, issues prevalent in traditional centralized management systems. Given the high risks involved, balancing data confidentiality with regulatory transparency is imperative. To overcome these limitations, a prototype system integrating blockchain technology and the Internet of Things (IoT) is proposed, featuring a multi-tiered consortium chain architecture. This system utilizes IoT sensors for real-time data collection, which is immutably recorded on the blockchain, while a hierarchical data structure (operational, supervisory, and public layers) manages access for diverse stakeholders. The results demonstrate that this approach significantly enhances data immutability, enables real-time multi-sensor data integration, improves decentralized transparency, and increases resilience compared to traditional systems. Ultimately, this blockchain-IoT framework improves the safety, transparency, and efficiency of spent fuel transportation, effectively resolving the conflict between confidentiality and transparency in nuclear data management and offering significant practical implications.}

\section{Introduction}

Spent nuclear fuel (SNF), also referred to as spent fuel, is the nuclear fuel that has been irradiated in and discharged from a nuclear reactor after its useful lifetime in power generation. It contains significant quantities of highly radioactive isotopes and requires careful management and secure transportation. Globally, thousands of shipments of radioactive materials, including both waste products and spent nuclear fuel, are transported daily.

The International Atomic Energy Agency (IAEA) issued the Regulations for the Safe Transport of Radioactive Material as early as 2012. Member states are required to transport radioactive materials, including spent nuclear fuel, in compliance with these regulations. With the implementation of these safety transport regulations, provided they are strictly followed, it can be assumed that all radioactive materials have been transported safely and reliably from their point of origin to their final destination. However, the transportation of radioactive materials continues to be one of the most vulnerable aspects of nuclear safety. Historical records indicate that spent nuclear fuel has been involved in numerous incidents, including theft, unauthorized disposal, and loss. According to the IAEA Incident and Trafficking Database (ITDB),\cite{IncidentTraffickingDatabase2019}Between 1993 and 2024, a total of 353 incidents involving the trafficking or malicious use of nuclear materials were recorded. Of these, approximately 86\% were related to trafficking activities, around 13\% involved fraud, and less than 2\% pertained to malicious use. Nuclear waste contains precious extractable metals, such as platinum and palladium, which are very expensive on the black market. In addition, certain radioactive materials, such as plutonium and uranium, can be utilized in the production of radiological weapons. If mishandled or leaked, these materials could lead to significant public concern and panic.\cite{huntDETERRINGNUCLEARRADIOLOGICAL}  The reasons for these vulnerabilities can be attributed to the following factors. First, nuclear waste is susceptible to loss or theft during transport in barrels to storage facilities because of inadequate protection measures. Second, there are significant management lapses in abandoned facilities. Third, insufficient international collaboration and weak regulatory systems in some countries hinder the effective tracking of radioactive sources throughout their entire lifecycle. Consequently, developing a comprehensive, secure, and stringent tracking and handling system for spent nuclear fuel has become an urgent priority for all nations.

This article aims to explore potential solutions for nuclear spent fuel accidents from the perspective of transportation tracking, utilizing distributed ledger technology (blockchain) to overcome the limitations of traditional data interaction methods. Additionally, it proposes a multi-layer data on-chain mechanism based on consortium chains to address the tension between the confidentiality of nuclear industry data (e.g., radiation dose and container location) and regulatory transparency (i.e., the disclosure by regulatory authorities to the public regarding regulatory activities, enforcement, and compliance of regulated entities).\cite{duRegulatoryTransparencyCitizen2023}For instance, the Nuclear and Industrial Safety Agency (NISA) and the Japan Nuclear Energy Safety Organization (JNES) examined the challenge of balancing confidentiality and transparency in nuclear data during the "Transparency of Nuclear Regulatory Activities" symposium. Their approach to addressing this issue involved fostering a balance between disclosing information and protecting proprietary secrets, encouraging increased public engagement, and establishing a robust trust framework.\cite{oecdTransparencyNuclearRegulatory2007a}However, in resolving the conflict between data confidentiality in the nuclear industry and regulatory transparency, this method faces challenges such as complex information categorization, delayed updates, vulnerability to tampering, difficulty in trust establishment, inefficient data sharing, weak traceability, restricted public involvement, and limited international cooperation. These challenges align precisely with the strengths of consortium chains. Meanwhile, reference can be made to Finland's SLAFKA blockchain-based data integrity management solution.\cite{cindyvestergaardSLAFKADemonstratingPotential2020}Propose a hierarchical management system.

%%%%%%%%%%%%%%%%%%%%%%%%%%%%%%%%%%%%%%%%%%
\section{Background}
\subsection{Characteristics of spent nuclear fuel}
As previously noted, spent fuel contains significant quantities of radioactive elements and requires appropriate management. During the cooling phase, spent nuclear fuel continuously emits decay heat due to the radioactive decay of nuclides. Excessive heat decay may compromise the functionality of the cooling systems in transportation equipment (e.g., dry storage casks), thereby increasing the risk of overheating. Consequently, continuous monitoring of temperature data for transportation equipment is essential during transit. Additionally, these nuclides exhibit high radioactivity levels. While radioactivity diminishes over extended cooling periods, certain long-lived radioactive nuclides in spent fuel (such as 238Pu, 241Am, and 137Cs) necessitate cooling durations exceeding a century.\cite{calicSpentFuelCharacterization2022}

There are two primary approaches for spent fuel management: one involves recycling the spent fuel by extracting reusable nuclear materials and repurposing them. Using chemical separation techniques, such as the PUREX process, uranium and plutonium can be recovered from spent fuel to produce mixed oxide (MOX) fuel, which effectively reduces the volume of radioactive waste.\cite{natarajanReprocessingSpentNuclear2017}The second option is underground storage of spent fuel. After being unloaded from the reactor, the spent fuel is cooled first, packaged and transported, and then directly buried underground as waste. Whether through the recycling and utilization of spent fuel or underground storage, the transportation of spent fuel must be based on the principle of safety.

\begin{figure}[h]
%\isPreprints{\centering}{} % Only used for preprints
\centering
\includegraphics[width=10 cm]{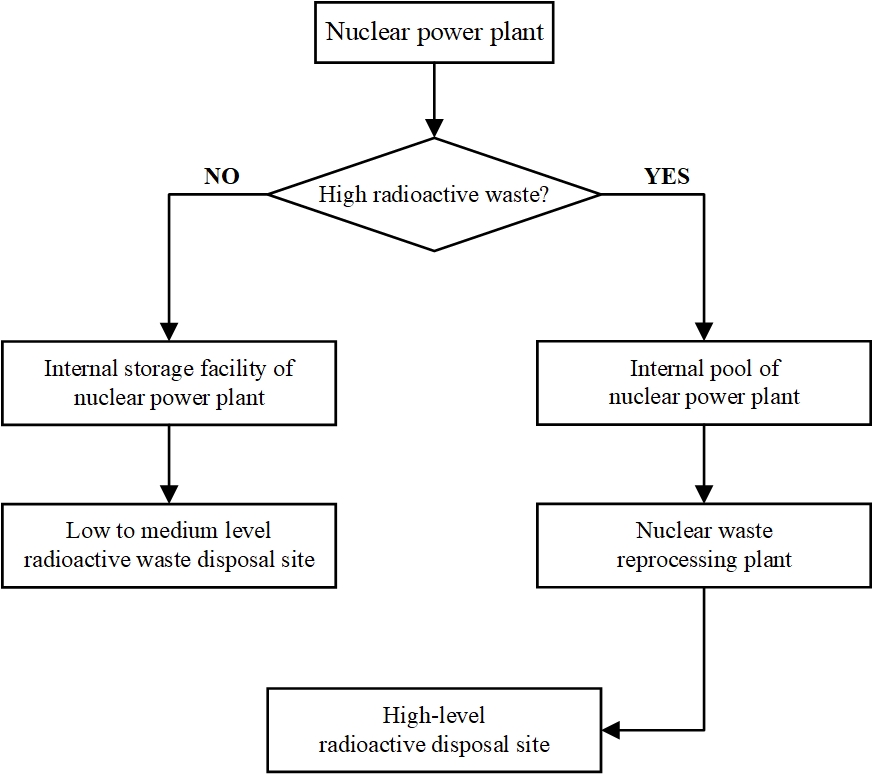}
\caption{The process of spent nuclear fuel treatment\label{fig1}}
\end{figure}  

Owing to the unique characteristics and high risks associated with spent nuclear fuel, its information management approach differs significantly from that of other materials. The information management strategy for spent nuclear fuel exhibits distinct features, necessitating transportation design specifically adapted to these attributes. Most importantly, continuous monitoring of radioactive containers during spent nuclear fuel transportation is critical to ensure their safety over extended periods (spanning years or even decades). Throughout its lifecycle, spent nuclear fuel is typically transported across long distances between various facilities, requiring continuous detection and analysis of its safety levels based on sensor data during transit. Any unforeseen incidents must be promptly reported. Additionally, the management of spent nuclear fuel transportation must establish mechanisms for providing appropriate access control for audits, thereby ensuring regulatory transparency.\cite{yessenbayevCombiningBlockchainIoT2024}

\subsection{Current management approach and its limitations}
Globally, the management approaches for spent nuclear fuel transportation are largely consistent. Taking China as an example, the country enforces a stringent application process for spent fuel transportation. Prior to transportation, the consignor is required to submit an application for a spent fuel transportation permit to the national nuclear safety regulatory authority, accompanied by detailed transportation plans, radiation monitoring reports, and other relevant data. Transportation can only commence upon approval. During transit, real-time data regarding the spent fuel's location, temperature, and radiation levels are transmitted to the consignor, carrier, and regulatory authorities via IoT-enabled devices installed on transportation vehicles or vessels, such as GPS positioning systems and Geiger counters. This facilitates continuous monitoring of the transportation process. In the event of any anomalies during transportation—such as excessive radiation levels or damage to the transportation container—the onboard emergency communication systems will immediately alert relevant emergency response agencies and simultaneously transmit real-time data from the accident site (e.g., video footage and images) to the emergency command center, enabling prompt emergency actions. Upon completion of transportation, the consignor, carrier, and consignee must consolidate and archive all data collected during the transportation process and report the relevant information to the national nuclear safety regulatory authority.\cite{yessenbayevCombiningBlockchainIoT2024}

Taking China as an example, the current management approach for nuclear spent fuel transportation in China, despite its relatively stringent process, is fundamentally based on a client-server architecture for nuclear waste management. This traditional client-server architecture exhibits limitations in processing and storing large volumes of real-time data, resulting in inefficiencies in data sharing and management. Specifically, data is stored locally and accessed only when necessary, precluding real-time interaction across the system. Given the numerous stakeholders involved in the transportation process, ensuring safe and accurate transportation requires strict reliance on point-to-point approvals and precise system-wide data interaction. Within this framework, authority is centralized among a few entities (primarily regulators), granting them primary control over the entire system. Similar deficiencies are evident in nuclear spent fuel transportation management systems in other countries, including issues such as insufficient real-time data accuracy, lack of transparency, and erosion of public trust. For instance, the 2014 accident at the Waste Isolation Pilot Plant (WIPP) in the United States, which was caused by incorrect container labeling and improper chemical filling, resulted in a chemical explosion that incurred substantial economic losses.

\section{Applying blockchain to nuclear waste tracking}
\subsection{Blockchain Characteristics}
In 2008, Satoshi Nakamoto published "Bitcoin: A Peer-to-Peer Electronic Cash System," marking the inception of Bitcoin. The underlying technology of Bitcoin, blockchain, has garnered significant attention due to its immutability, openness, and other distinctive features. At its core, blockchain is a distributed ledger system. Unlike traditional centralized ledgers (characterized by centralized transactions), blockchain does not rely on a single centralized storage entity. In a blockchain network, each participant (node) maintains a complete copy of the ledger. Blockchain consists of a series of blocks, where each block contains a specific number of transaction records. These records may include currency transfers (such as Bitcoin transactions), smart contract execution logs, and various other types of data. Blocks are interconnected through cryptographic techniques, primarily hash algorithms. Each new block incorporates the hash value of the preceding block, forming a chain-like structure that ensures the immutability of the entire blockchain.

In addition to its immutability, blockchain also exhibits openness. Owing to this characteristic, various industries have shown keen interest in adopting blockchain technology, particularly consortium chains. However, in practical applications, organizations must balance data sharing with the protection of internal data while ensuring compliance with regulatory requirements. This necessitates addressing the challenge of enabling specific user groups within different organizations to access data at varying sensitivity levels appropriately. Failure to do so may result in the risk of data privacy breaches.\cite{YaoZhongJiangGuanYuQuKuaiLianYuanLiJiYingYongDeZongShu2017}

Blockchain can be categorized into three primary types: public chain, consortium chain, and private chain. Among these, the public chain enables anyone to participate in both the maintenance and reading of blockchain data, operates without control by any single institution, and ensures open and transparent data. However, it is characterized by relatively low transaction performance and a slower overall development pace, as exemplified by Bitcoin. The consortium chain, leveraging its specific organizational structure and consensus mechanism, enhances system performance and security while maintaining a certain degree of decentralization. Notable examples include Hyperledger Fabric and Openchain. The private chain offers a high level of data privacy protection but exhibits relatively limited openness.Table \ref{tab:blockchain_comparison} provides a summary of the distinctions among public chain, consortium chain, and private chain across various dimensions, including access permissions, number of nodes, consensus mechanisms, and data privacy.

\begin{table}[htbp] % 'h'ere, 't'op, 'b'ottom, 'p'age placement preference
    \centering % Center the table horizontally
    \caption{Blockchain Comparison} % Add the caption
    \label{tab:blockchain_comparison} % Add a label for cross-referencing
    \begin{tabular}{@{}lccc@{}} % Define column alignment: l=left, c=center. @{} removes extra space at edges.
        \toprule % Top rule from booktabs
        % Use \multirow{2}{*} to make "Feature" span two rows vertically. '*' calculates natural width.
        \multirow{2}{*}{\textbf{Feature}} & \textbf{Public} & \textbf{Consortium} & \textbf{Private} \\
         & \textbf{Blockchain} & \textbf{Blockchain} & \textbf{Blockchain} \\
        \midrule % Middle rule from booktabs
        \textbf{Access Permissions} & Open & Requires permission & Private \\
        \textbf{Nodes Number} & Many & Fewer & Very few \\
        \textbf{Consensus Mechanism} & PoW, PoS, etc. & PBFT, Raft, etc. & Customized \\
        \textbf{Data Privacy} & Transparent & Semi-transparent & Private \\
        \textbf{Regulatory} & Generally unregulated & Meet the regulatory & Meet the regulatory \\
        \bottomrule % Bottom rule from booktabs
    \end{tabular}
\end{table}

\subsection{Blockchain Characteristics}
\subsubsection{Application of Blockchain in the Supply Chain}
The unique attributes and advantages of blockchain technology can substantially enhance supply chain management. By increasing the transparency of tracking goods movement across the entire supply chain, blockchain is expected to significantly improve supply chain operations. This enhancement will lead to higher data accuracy and immutability, thereby reducing instances of fraud and minimizing errors. Furthermore, the automation of certain processes and the reduction in reliance on intermediaries will boost overall supply chain efficiency. Additionally, the unidirectional flow of information, which automatically encrypts all communications, enhances security when sensitive information is shared among parties.\cite{sharabatiBlockchainTechnologyImplementation2024}Since blockchain transaction records are more accessible for querying, they exhibit enhanced auditability and transparency. In contrast to multi-layered approval processes, blockchain systems are more efficient and easier to manage.

Blockchain demonstrates greater applicability in the supply chain management of high-value and hazardous materials. Its decentralized and transparent nature plays a critical role in preventing theft, counterfeiting, and ensuring the authenticity of goods. Recently, Yi Dai et al. introduced a blockchain-based access control system tailored for the dangerous goods supply chain. By leveraging blockchain technology and smart contracts, this system offers a secure, efficient, and transparent access control solution for the supply chain of hazardous materials, thereby significantly enhancing the efficiency and security of supply chain operations.\cite{daiBlockchainBasedAccessControl2024a} A substantial body of research supports the adoption of blockchain in high-value and hazardous material supply chains, thereby establishing a robust foundation for its application in the specific context of nuclear waste tracking. The key advantages often highlighted in scholarly literature—enhanced security, transparency, and traceability—are intrinsically aligned with the stringent requirements associated with nuclear waste management.

\subsubsection{Current research progress}
SLAFKA represents the world's first blockchain prototype explicitly tailored for nuclear safeguards, collaboratively developed by the Finnish Radiation and Nuclear Safety Authority (STUK), the Stimson Center, and the University of New South Wales (UNSW). This system leverages DLT to streamline the reporting and inspection processes of nuclear materials, addressing challenges related to data integrity, cyber threats, and inefficiencies inherent in traditional reporting systems. By establishing a shared, immutable ledger, SLAFKA facilitates secure and transparent communication of safeguard data between operators and regulatory authorities. The prototype's design adheres to the requirements of Finland's current State System for Accounting and Control of Nuclear Material (SSAC) and aligns with the regulations set forth by the European Commission. Technically, SLAFKA is constructed on the Hyperledger Fabric platform, functioning as a private, permissioned blockchain. This architecture inherently satisfies the stringent demands for data confidentiality and access control in nuclear safeguards information management. Within SLAFKA's simulation environment, it encompasses three nuclear facility operators—two nuclear power plants and one deep geological repository—and three tiers of regulatory authorities representing national, regional, and international levels. The system abstracts actual nuclear material batches into digital assets with unique identifiers ("batches") for management and circulation within the distributed ledger. Functionally, SLAFKA successfully enables the input, storage, display, and output of nuclear material accounting information, fulfilling the reporting requirements of institutions such as the International Atomic Energy Agency (IAEA) and the European Atomic Energy Community (Euratom), including formats like ICR, PIL, and MBR. SLAFKA supports a comprehensive range of key nuclear material management transactions, encompassing material production, domestic and international transportation and receipt, nuclear conversion and loss, as well as changes in attributes or location and re-batching. These transactions meticulously document detailed batch attributes, including material balance areas (MBA), key measurement points (KMP), material form, container type, weight, and isotopic composition, among other critical parameters.\cite{cindyvestergaardSLAFKADemonstratingPotential2020}

Despite its significant achievements, it is crucial to acknowledge that SLAFKA remains a proof-of-concept prototype at present. It relies on fictional data and has not been fully security-hardened for production environments. Additionally, it does not yet leverage all the advanced capabilities offered by Hyperledger Fabric, such as channels, private data collections, and endorsement policies. Its current scope primarily centers on fundamental nuclear material accounting and does not yet encompass the specific requirements of Additional Protocol (AP) reporting or the complex provisions of multinational nuclear cooperation agreements (NCAs).

A few years ago, the Sellafield DLT Field Lab in the UK collaborated with RKVST to investigate the potential of distributed ledger technology (DLT) for digitizing the tracking of nuclear waste data, leading to the establishment of the Sellafield DLT Field Lab project. As the largest nuclear decommissioning organization in the UK, Sellafield faced significant challenges in nuclear waste management and worker qualification management, necessitating improvements in productivity and cost reduction through digital transformation. The project was executed in two phases: the first phase identified two critical areas—nuclear waste tracking and worker qualification management—and engaged technical partners; the second phase involved RKVST and Condatis in developing prototype solutions. The project successfully enabled transparent tracking of nuclear waste, rapid verification of worker qualifications, minimized manual processes, and strengthened data security and audit capabilities.\cite{HarnessingPowerDistributed}

These studies still offer opportunities for improvement in critical areas such as data integrity and network threats and have yet to fully address the specific requirements of complex cross-border nuclear cooperation. Building on the aforementioned research advancements, this paper aims to further investigate the potential applications of blockchain technology in tracking the transportation of spent nuclear fuel.

\section{System Design}
\subsection{Integrated System of Blockchain and IoT}
Designing a secure spent nuclear fuel tracking system requires the implementation of robust data collection technology. The Internet of Things (IoT) is particularly well-suited for integration into such systems due to its ability to facilitate real-time data collection. Unlike traditional computers, IoT devices are characterized by their extensive coverage despite having limited computational power. Numerous studies have explored the application of blockchain technology within the Internet of Things ecosystem. For instance, Uddin et al. conducted a comprehensive review several years ago that examined the applications, challenges, and solutions associated with integrating blockchain technology into IoT systems. Their work highlighted key challenges, such as resource constraints, data processing limitations, and privacy concerns, while also proposing viable solutions to address these issues.\cite{uddinSurveyAdoptionBlockchain2021}These studies provide a solid foundation for the integration of IoT with blockchain technology.

\subsubsection{Selection of IoT Sensors}
The type of IoT sensors selected determines the breadth and quality of data that can be collected. When monitoring the transportation of dangerous or sensitive materials, it is essential to select a diverse set of IoT sensors to ensure the effective collection of real-time data for integration into the blockchain. Below are the types of sensors that may be applicable:

\begin{itemize} % Use the enumerate environment for numbered lists
    \item \textbf{GPS trackers:} Provide real-time geographical location data of transportation vehicles. For instance, the U.S. Department of Energy’s Transportation Tracking and Communication System (TRANSCOM) uses satellite tracking to frequently provide location updates, once per minute.\cite{craigARGUSRFIDMONITORING2013}

    \item \textbf{Radiation detectors:} Ensure the integrity of spent nuclear fuel, monitor external radiation dose rates, and detect violations. For instance, by surrounding spent fuel barrels with SiLiF and SciFi sensor arrays, the status of spent fuel can be detected in real time.\cite{cosentinoSiLiFNeutronCounters2021}

    \item \textbf{Temperature sensor:} Monitors the environmental conditions and the thermal state of the packaging during transportation.

    \item \textbf{Shock/vibration sensor:} Detect physical impacts and drops during transportation.

    \item \textbf{Anti-tampering sealed sensor:} Prevents unauthorized damage and disassembly of spent fuel packaging.

    \item \textbf{RFID tags:} Radio Frequency Identification tags, which can be used for automatic identification and tracking of sensor data. They can enhance security, safety and service (3S) while providing real-time status and environmental data.
\end{itemize}

\subsubsection{Comparative Analysis: IoT-Blockchain vs. Traditional Tracking Methods}
Compared to traditional client-server tracking methods, which typically depend on centralized databases, manual reporting, and less precise monitoring, the integration of IoT and blockchain provides a shared distributed ledger. This ledger serves as a single source of truth and enables near real-time access for all authorized stakeholders, such as regulators, shippers, consignees, and emergency responders. As a result, spent fuel becomes fully transparent to regulators throughout its entire transportation lifecycle, effectively breaking down the information silos prevalent in traditional systems where data is often fragmented across multiple platforms.\cite{ecemisExploringBlockchainNuclear2024}The integration of the Internet of Things (IoT) and blockchain offers substantial advantages in auditability and traceability. Each transaction or data entry recorded on the blockchain is timestamped and cryptographically linked, creating a permanent, verifiable, and auditable event tracking record. This capability supports regulatory compliance checks and accident investigations while enabling end-to-end traceability of nuclear waste packaging from origin to final disposal. It effectively addresses limitations of traditional methods, such as manual logs, decentralized databases, or records susceptible to tampering or loss. In terms of efficiency and cost, this system automates data recording and verification (potentially through smart contracts), reduces dependency on intermediaries, and streamlines reporting processes. Moreover, real-time data availability enhances decision-making speed and quality. Studies on general supply chains demonstrate that blockchain technology can significantly lower management costs.

Take the TRANSCOM Communication Center of the U.S. Department of Energy (DOE) as an example. TRANSCOM is an unclassified tracking and communication web application of the DOE, used to monitor the progress of special nuclear materials (domestic), foreign nuclear materials, transuranic waste, and any other authorized goods. Authorized TRANSCOM users (such as DOE shippers, carriers, state and local governments, and various federal agencies) can access the web application from their computers or any mobile devices. TRANSCOM is the traditional tracking system based on client-server mentioned above.Table \ref{tab:transcom_vs_iot} provides a comparison of key functions between the hypothetical integrated IoT-Blockchain system and the traditional tracking system TRANSCOM:

\begin{table}[htbp] % Placement specifier (here, top, bottom, page)
    \centering % Center the table
    \caption{TRANSCOM vs. IoT-Blockchain system} % Table caption
    \label{tab:transcom_vs_iot} % Label for cross-referencing

    % Define a new column type 'L' which is like 'X' (variable width, wrapping)
    % but left-aligned (\raggedright) instead of justified.
    \newcolumntype{L}{>{\raggedright\arraybackslash}X}

    % Use tabularx to make the table fit the text width (\textwidth).
    % Column 1: Fixed width (p{3cm}), adjust '3cm' as needed for your layout. Left-aligned.
    % Column 2 & 3: Use the new 'L' type for auto-wrapping, left-aligned text.
    \begin{tabularx}{\textwidth}{@{} p{3.2cm} L L @{}} % @{} removes padding at edges
        \toprule
        \textbf{Feature} & \textbf{TRANSCOM} & \textbf{IoT-Blockchain system} \\
        \midrule
        \textbf{Data Immutability}
        & Rely on conventional database security, access controls, audit logs.
        & Cryptographically enforced via hashing and consensus; tamper evident. \\
        \addlinespace % Add a small vertical space between rows (from booktabs)

        \textbf{Real-time Multi-Sensor Data}
        & Primarily GPS location \& messaging; sensor integration via add-ons.
        & Designed for native integration of diverse real-time sensor data streams. \\
        \addlinespace

        \textbf{Decentralization}
        & Typically centralized architecture.
        & Distributed ledger replicated across multiple authorized nodes. \\
        \addlinespace

        \textbf{Stakeholder Transparency}
        & Access controlled by central authority.
        & Shared ledger accessible by all permissioned stakeholders; configurable transparency. \\
        \addlinespace

        \textbf{Audit Trail Verifiability}
        & Relies on database logs; potentially alterable by administrators.
        & Permanent, time-stamped, cryptographically verifiable transaction history. \\
        \addlinespace

        \textbf{Long-term Integrity Assurance}
        & Dependent on database longevity, backups, institutional continuity.
        & Enhanced resilience via distribution and cryptographic linking. \\
        \addlinespace

        \textbf{Resistance to Single Point Failure}
        & Central server/database represents a potential single point of failure.
        & Distributed nature eliminates single points of failure for data records. \\
        \addlinespace

        \textbf{Trust Mechanism}
        & Trust in the central authority (DOE) managing the system.
        & Through shared, immutable data and consensus protocols. \\
        \bottomrule
    \end{tabularx}
\end{table}

\subsubsection{The architecture of the IoT-Blockchain system}

\begin{figure}[h]
\centering
\includegraphics[width=13 cm]{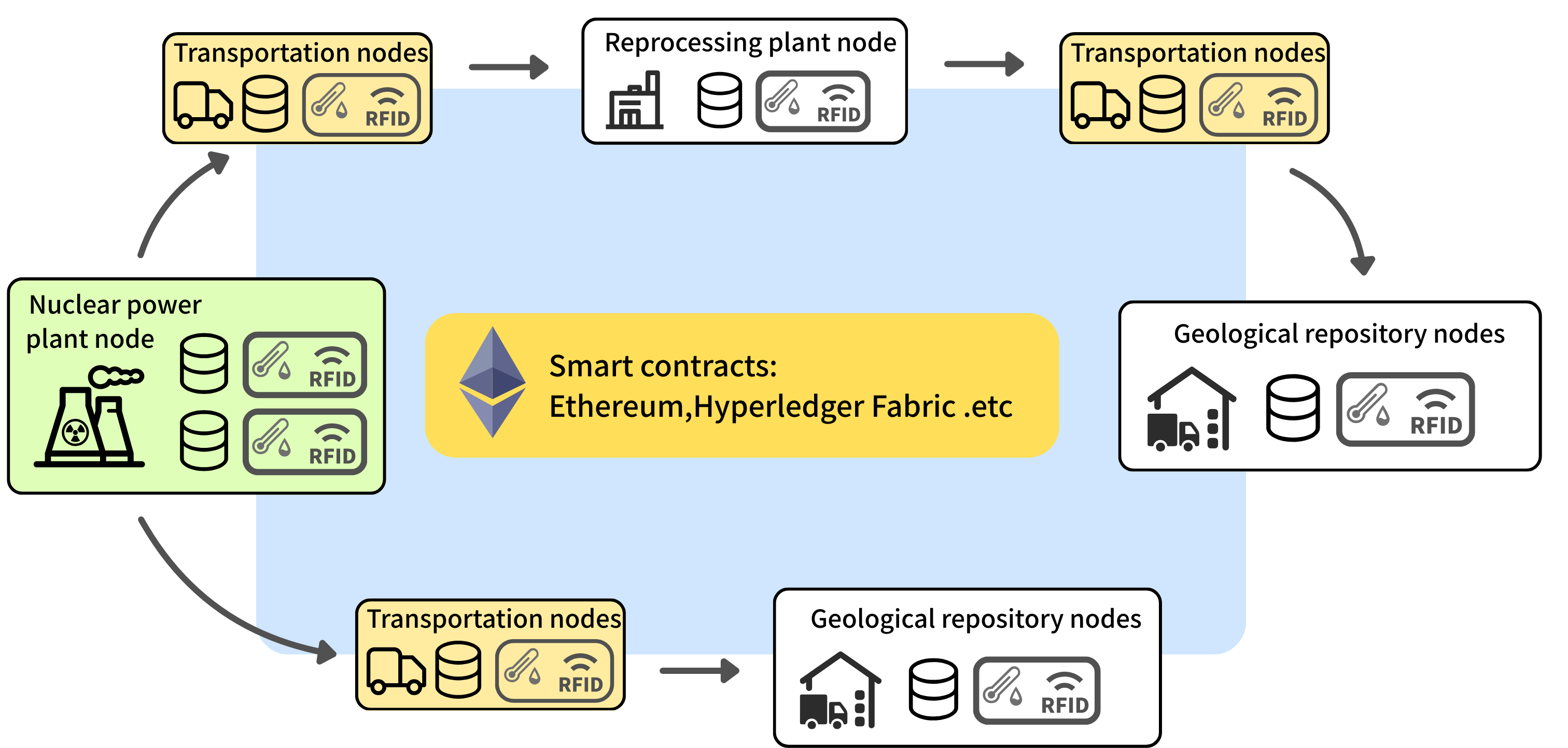}
\caption{The process of spent nuclear fuel treatment\label{fig2}}
\end{figure}

Figure \ref{fig2} illustrates a nuclear spent fuel tracking system that integrates blockchain technology with the Internet of Things (IoT), enhancing the transparency and security of nuclear spent fuel management through dynamic monitoring and data sharing. The system's foundation is a blockchain-based smart contract platform, such as Ethereum or Hyperledger Fabric, which collaborates with IoT devices (e.g., RFID tags) to ensure comprehensive oversight throughout the entire lifecycle of nuclear spent fuel. Within this framework, IoT devices (as indicated by the RFID box in the figure) serve a critical function. These devices are strategically placed at key locations, including nuclear power plants, transportation hubs, reprocessing facilities, and geological storage sites, to continuously monitor the status of nuclear spent fuel containers. This includes core parameters such as temperature, humidity, radiation levels, and GPS location. Communication between IoT devices occurs via low-power wide-area networks, such as LoRa/LoRaWAN or LTE.\cite{khutsoaneIoTDevicesApplications2017}The collected data is uploaded to the blockchain system, ensuring its real-time availability and accuracy. IoT devices not only enable dynamic monitoring but also enhance the transparency of multi-party collaboration. For example, during transportation, IoT sensors continuously gather container status data and synchronize it in real time to the blockchain. This allows transportation companies, regulatory authorities, and destination facilities to simultaneously access the latest status information. Such an information sharing mechanism ensures that all relevant parties maintain real-time awareness of the spent nuclear fuel's status, thereby strengthening the transparency and trustworthiness of the management process.\cite{yessenbayevCombiningBlockchainIoT2024} Once the data collected by Internet of Things (IoT) sensors is uploaded to the blockchain, it becomes permanently recorded and tamper-proof. This permanence ensures the authenticity and integrity of the information while providing verifiable public records for regulatory agencies and the public. For instance, when a geological storage node receives spent nuclear fuel, IoT devices verify whether the container status complies with safety standards and record the results on the blockchain, making them accessible for review by all relevant parties.

To effectively manage the system's complexity and ensure its modularity and scalability, this architectural design employs a layered architecture pattern. This pattern is widely utilized in the development of Internet of Things systems, with its primary advantage being the clear delineation of each component's functional scope and the facilitation of standardized interaction interfaces between layers.[18] Based on the general Internet of Things architecture model\cite{oktianHierarchicalMultiBlockchainArchitecture2020}.As shown in Figure \ref{fig3}, the architecture of this system is divided into five layers:

\begin{figure}[h]
\centering
\includegraphics[width=6 cm]{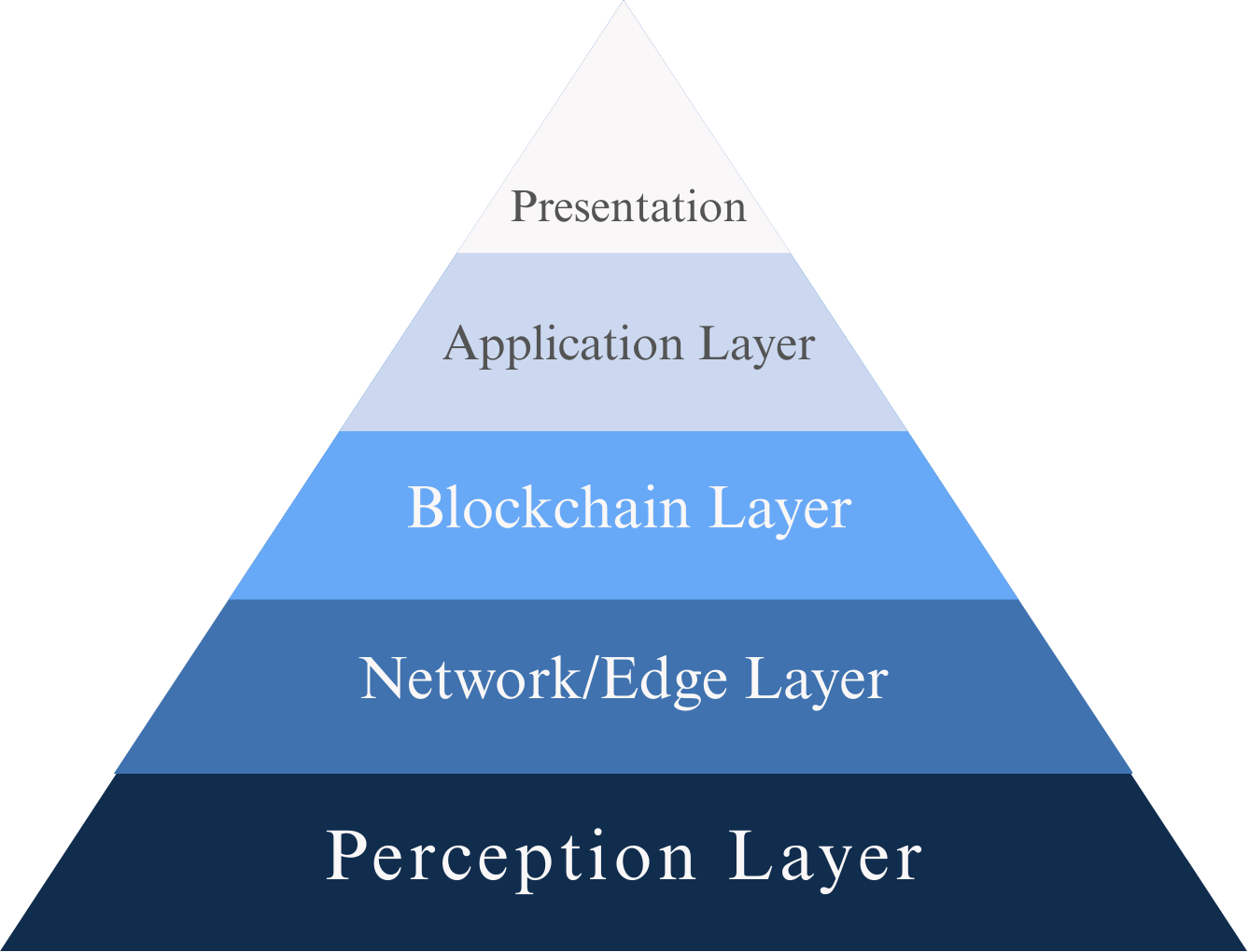}
\caption{System hierarchical architecture\label{fig3}}
\end{figure}

\begin{enumerate} % Use the enumerate environment for numbered lists
    \item \textbf{Perception Layer:} This layer consists of various Internet of Things (IoT) physical sensors (e.g., temperature, radiation, RFID, etc.) installed on nuclear waste transport containers or vehicles. These sensors are capable of collecting raw data in real time related to the container's location, environmental conditions, physical status, and identification information.

    \item \textbf{Network/Edge Layer:} This layer is responsible for securely and reliably transmitting the data collected by the perception layer to the subsequent processing layers. It supports a wide range of communication protocols (both short-range and long-range) and network technologies (e.g., cellular networks, satellite communications). Additionally, it incorporates edge computing, a paradigm that performs computation at the network edge. By preprocessing, filtering, and aggregating data on gateway devices near the data source, edge computing reduces the volume of transmitted data, minimizes latency, and improves system responsiveness.\cite{caoOverviewEdgeComputing2020}

    \item \textbf{Blockchain Layer:} This layer acts as the system's core trust layer, leveraging distributed ledger technology (specifically Hyperledger Fabric) to securely and immutably record verified transportation data. Moreover, smart contracts are deployed in this layer to execute predefined business logic, including data validation, status updates, access control enforcement, and alert triggering mechanisms.

    \item \textbf{Application Layer:} Serving as an intermediate layer, this layer connects the blockchain layer with the presentation layer. It retrieves data from the blockchain ledger and processes it into formats that are accessible to users or external systems. Additionally, it provides API interfaces to enable seamless integration with external systems, including regulatory reporting systems and emergency response systems.

    \item \textbf{Presentation Layer:} This layer serves as the interface through which users interact with the system. It delivers customized dashboards and advanced visualization tools tailored to the specific needs of different user roles, including shippers, carriers, regulatory agencies, and emergency response personnel. Examples of these tools include sensor data charts, historical audit trails, and real-time alert notifications.
\end{enumerate}

\subsection{Consortium Chain: Balancing Confidentiality and Regulatory Transparency}
\subsubsection{Problem Overview}
The preceding analysis has underscored the intrinsic conflict between confidentiality and transparency in the management of nuclear data. Throughout the entire lifecycle of spent nuclear fuel—encompassing generation, transportation, storage, and final disposal—this material presents considerable risks that necessitate rigorous confidentiality protocols. The transportation of nuclear waste entails the handling of highly sensitive information, including the precise locations of containers, radiation dose levels, detailed transport routes, and security measures. Unauthorized disclosure of such information could result in significant security threats, including theft, sabotage, or potential terrorist activities. To address these risks, nations implement stringent access controls to critical information. For example, the Office for Nuclear Regulation in the United Kingdom requires carriers to submit detailed Transport Security Statements and ensures that only individuals who have undergone thorough security vetting are permitted access to sensitive nuclear data. Similarly, the United States Nuclear Regulatory Commission (NRC) enforces strict protective measures for confidential information, Safeguards Information, and Sensitive Unclassified Non-Safeguards Information.\cite{InformationSecurity}

Concurrently, there has been a notable demand for transparency in nuclear waste management practices, particularly in the context of transportation, from both the public and regulatory authorities. The historical culture of secrecy within the nuclear sector, combined with concerns regarding environmental contamination, has significantly undermined public trust. In order to restore and maintain this trust, regulatory agencies are required to provide comprehensive disclosures regarding their oversight activities, the outcomes of compliance verifications, and the adherence of regulated entities to established regulatory standards. For instance, the U.S. Department of Energy (DOE) has faced criticism for its lack of transparency in the management of certain low-level waste disposal strategies, primarily due to its failure to effectively communicate the rationale behind its decisions. Enhancing transparency is deemed crucial for improving the decision-making process and cultivating trust among various stakeholders, including local communities, state regulatory bodies, and tribal nations.\cite{oecdTransparencyNuclearRegulatory2007a}Nevertheless, an inherent and difficult-to-reconcile tension exists between the imperative for regulatory transparency and the necessity to ensure the security and confidentiality of sensitive transportation data. Over-disclosure could compromise transportation safety, whereas over-securing such information might exacerbate public skepticism and erode trust.

\subsubsection{Demand for data management}
In order to effectively establish a robust tracking system for the transportation of nuclear waste, it is imperative to undertake a comprehensive examination of the unique and frequently conflicting data management requirements associated with this field. These requirements include the implementation of rigorous confidentiality protocols for highly sensitive information, alongside the need for regulatory compliance and transparency.

The subsequent section delineates the highly sensitive information pertinent to the nuclear waste transportation process, which is governed by the confidentiality protocols of the system.

\begin{enumerate}[leftmargin=*] % Use enumerate for numbered lists. leftmargin=* aligns items with paragraph margin.
    \item \textbf{Real-time container location:} Provide precise GPS coordinate data to prevent potential interception or attacks during transportation.

    \item \textbf{Radiation measurement:} Monitor real-time or historical radiation levels inside and outside the container, which may indicate material properties or container integrity.

    \item \textbf{Transportation route and schedule:} It is essential to provide comprehensive details regarding planned routes, alternative routes, departure times, and estimated arrival times. This information is vital for security planning purposes, and any unauthorized dissemination of such data could substantially elevate associated risks.

    \item \textbf{Security details:} Specify security systems, procedures, personnel deployment, and emergency response plans outlined in the Transport Security Statement (TSS).

    \item \textbf{Personnel information:} Include identities, security clearance levels, and specific responsibilities of individuals involved in transportation operations.

    \item \textbf{National security information:} This may include classified information, restricted data, protective measures, or sensitive yet unclassified information that is intricately associated with the physical security of nuclear materials or facilities.

    \item \textbf{Handover points and procedures:} Sensitive operational details such as specific handover locations, times, and confirmation protocols for transferring responsibility.
\end{enumerate}

In the domain of nuclear waste management, there exists a considerable necessity for regulatory transparency. Attaining this transparency is crucial for cultivating and maintaining public trust, in addition to ensuring accountability. On December 7, 1993, the U.S. Department of Energy (DoE) introduced its "Openness Initiative," which aimed to bolster public confidence by augmenting transparency and disseminating extensive information regarding all departmental activities. This initiative set a benchmark for the disclosure of nuclear-related data and information, thereby incorporating the public as active stakeholders in the regulatory process.\cite{schaperLookingDemarcationNuclear2004}Regulatory authorities are expected to disclose specific types of information to the public, local communities, state governments, tribal nations and other stakeholders. This typically includes: regulatory activities and processes, compliance status, regulatory enforcement actions, anonymous data, etc. According to the latest publication of the International Atomic Energy Agency on the safety of the transport of nuclear and other radioactive materials (2024)\cite{internationalatomicenergyagencySecurityNuclearOther2024}, Table \ref{tab:confidentiality_transparency_summarized} assesses the requirements for confidentiality and transparency in the transportation of nuclear waste:

\begin{table}[htbp] % Placement specifier (here, top, bottom, page)
    % Updated caption to reflect summarized content
    \caption{Requirements for confidentiality and transparency}
    % Using a new label for this version
    \label{tab:confidentiality_transparency_summarized}

    % Define a new column type 'L' which is like 'X' (variable width, wrapping)
    % but left-aligned (\raggedright) instead of justified.
    \newcolumntype{L}{>{\raggedright\arraybackslash}X}

    % Use tabularx to make the table fit the text width (\textwidth).
    % Column types: L (ragged X), c (centered), c (centered), L (ragged X)
    \begin{tabularx}{\textwidth}{@{} L c c L @{}} % @{} removes padding at edges
        \toprule
        % Header Row: Manually bold all headers; updated last header
        \textbf{Data Category} & \textbf{Confidentiality} & \textbf{Transparency} & \textbf{Conflict Points} \\
        \midrule
        \textbf{Physical Location/Path} & High & Low & Balance security monitoring with tracking needs. \\
        \addlinespace % Adds a small vertical space (from booktabs)

        \textbf{Radiation dosage} & High & Low & Prove compliance without revealing sensitive data patterns. \\
        \addlinespace

        \textbf{Security Programs} & High & Low & Show security plan approval without revealing vulnerabilities. \\
        \addlinespace

        \textbf{Personnel Identity} & High & Low & Verify personnel qualifications without exposing private info. \\
        \addlinespace

        \textbf{Interaction Records} & Medium & Low & Prove handover occurred without revealing sensitive operational details. \\
        \addlinespace

        \textbf{Compliance Audit Reports} & Medium & High & Issue credible audits while protecting sensitive operational data within them. \\
        \addlinespace

        \textbf{Regulatory Decision} & Low & High & Disclose decisions/processes while protecting confidential info/trade secrets. \\
        \addlinespace

        \textbf{Aggregated Operational Statistics} & Low & High & Ensure aggregated data is reliable and free of sensitive details. \\
        \addlinespace

        \textbf{Environmental Impact Statements} & Low & High & Ensure disclosed EIS is complete, accurate, and decisions are justified. \\
        \bottomrule
    \end{tabularx}
 
\end{table}

The analysis of these conflict points highlights a critical insight: the solution cannot simply be "fully open" or "fully closed." The system design must be capable of distinguishing among different types of data, varying time dimensions (real-time versus historical/aggregate data), and diverse audiences (operators, regulators, and the public). This nuanced differentiation serves as the cornerstone of a multi-layered architecture design. Furthermore, merely granting data access rights is inadequate; in light of trust deficits, the system must also provide mechanisms to verify the integrity and authenticity of disclosed information, such as through tamper-proof records and cryptographic proofs—capabilities that represent the core strengths of blockchain technology.

\subsubsection{Multi-layer architecture based on consortium blockchain}
In order to develop a system that effectively reconciles the requirements of confidentiality and transparency in the tracking of nuclear waste, it is essential to implement particular blockchain technologies and architectural frameworks. This section will explore the core principles of consortium blockchains and multi-layer blockchain architectures, as well as assess their relevance and suitability for this specific application context. 

A consortium chain is a type of blockchain that lies between public and private chains, jointly managed and maintained by a pre-selected group of organizations. It combines the controllability of private chains with some of the decentralized features of public chains, making it suitable for scenarios where multiple parties collaborate but do not wish for complete openness. Consortium chains have been proven to have great potential in improving supply chain management\cite{chenSurveyConsortiumBlockchain2024}, its core features include:

\begin{itemize}[leftmargin=*]
    \item \textbf{Permissioned Access:} Only entities authorized by the consortium can join the network, access data, submit transactions, or participate in the consensus process.

    \item \textbf{Identity Management:} Enhance accountability through digital identity authentication; consortium chains generally mandate that participants possess known and verifiable identities. This is accomplished via member service providers (MSPs) and public key infrastructure (PKI). Each participant is assigned a digital identity to authenticate and authorize their operations on the network.

    \item \textbf{Controlled Transparency:} Data is private by default and access policies can be flexibly set.

    \item \textbf{Governance:} The rules and protocols are upgraded by the joint decision of the members.

    \item \textbf{High Performance and Efficiency:} By adopting efficient consensus mechanisms, such as voting-based Byzantine Fault Tolerance algorithms like PBFT and Raft, consortium chains generally achieve higher transaction throughput (TPS) and significantly reduced transaction confirmation latency compared to public chains.
\end{itemize}

The consortium chain model is highly aligned with the requirements for nuclear waste transportation tracking. The transportation process of spent nuclear fuel involves a well-defined set of participants, including waste producers, transportation companies, receiving facilities, and regulatory agencies. These entities must collaborate while remaining subject to strict control. The permissioned access and identity management mechanisms of consortium chains ensure that only vetted and legitimate entities can participate in the system, thereby preventing unauthorized access. Its controllable transparency allows core participants to share necessary operational information while isolating sensitive data from external environments. The shared control model fosters trust and cooperation among participants. In contrast to fully open public chains (which lack sufficient security) or private chains (which may lack multi-party trust) controlled by a single entity (which may fail to establish multi-party trust), consortium chains provide a more balanced and appropriate framework. By combining the permissioned access control of consortium chains with the data isolation capabilities of a multi-layer architecture, an effective technical solution is provided for addressing the core challenges in nuclear waste tracking. Consortium chains ensure participant legitimacy and enforce basic access control, while the multi-layer architecture enables more granular data partitioning and access management. Sensitive operational data can be securely isolated in the lowest layer, accessible only to core operators, while higher layers meet regulatory and public demands for verifiable transparency through cryptographic proofs (such as zero-knowledge proofs) or securely aggregated data.

Based on the aforementioned demand analysis and system stratification, this section focuses on the blockchain layer in the previously introduced stratified model, further subdividing it into three sub-layers: the operational layer, the supervisory layer, and the public layer. The operational layer represents the most strictly permission-controlled tier, tasked with recording and managing real-time, highly sensitive operational information during transportation, such as precise GPS location data, radiation dose measurements, detailed transportation route planning, and security records. Its primary objective is to ensure the physical security and operational efficiency of transportation, with access restricted to authorized personnel and security teams following the "need-to-know" principle. By utilizing technologies like Hyperledger Fabric's Private Data Collections (PDCs), this layer ensures that sensitive information is transmitted point-to-point exclusively among authorized nodes, remaining confidential from other consortium chain members (including ordering nodes). Only the hash values of the data are anchored on the main ledger for verification purposes.

The supervisory layer acts as a critical bridge between operational confidentiality and regulatory transparency, storing verified compliance information, regulatory approval records, validation records of key transportation events, and regulatory audit logs. This layer enables regulatory authorities and relevant compliance departments to access necessary information to fulfill their oversight responsibilities. To achieve this without exposing operational details, the supervisory layer employs privacy-enhancing technologies such as Zero-Knowledge Proofs (ZKPs), which can demonstrate compliance without revealing the underlying data.  

The public layer is designed to meet the transparency requirements of the public and various stakeholders, thereby promoting trust through openness. This layer comprises aggregated, anonymized, or desensitized data, including comprehensive transportation statistics, summary reports at a high level, links to publicly accessible regulatory documents, and hash values of data from the supervisory layer to authenticate public information. Access to this layer is generally public and restricted to read-only permissions, ensuring widespread accessibility while preserving traceability and verifiability through the use of cryptographic techniques.

Through this layered design, the system achieves precise management of information visibility and access rights based on the nature of the data and its intended audience, effectively balancing the high confidentiality required for secure nuclear material transportation with the regulatory transparency needed to build public trust.As shown in Table \ref{tab:blockchain_layering}:

\begin{table}[htbp] % Placement specifier (here, top, bottom, page)
    \centering % Center the table
    \caption{Specific layering of the blockchain layer} % Table caption
    \label{tab:blockchain_layering} % Label for cross-referencing

    % Define a new column type 'L' which is like 'X' (variable width, wrapping)
    % but left-aligned (\raggedright) instead of justified.
    \newcolumntype{L}{>{\raggedright\arraybackslash}X}

    % Use tabularx to make the table fit the text width (\textwidth).
    % Column 1: Bold text (>{\bfseries}), uses 'L' type (ragged-right, variable width).
    % Columns 2, 3, 4: Use 'L' type.
    \begin{tabularx}{\textwidth}{@{} >{\bfseries}L L L L @{}} % @{} removes padding at edges
        \toprule
        % Header Row: First column is bold by column type; others need \textbf
        Layer Level & \textbf{Purpose} & \textbf{Key Data Types} & \textbf{Access Restrictions} \\
        \midrule
        Operation Layer
        & Record real-time sensitive operation data to ensure transportation safety and efficiency
        & Raw sensor data, detailed routes, security logs, etc.
        & High degree of restriction, based on tasks and permissions, need "need to know" \\
        \addlinespace % Adds a small vertical space (from booktabs) for readability

        Supervision Layer
        & Store verification of compliance information, supervision activity records, etc.
        & Compliance certificates, supervision approval records, etc.
        & Permissioned access, based on roles and policies \\
        \addlinespace

        Public Layer
        & Provide publicly verifiable information to meet transparency and establish trust
        & Aggregated statistical data, anonymized compliance summaries, regulatory data layer hash
        & Wide read access, strict control over write permissions \\
        \bottomrule
    \end{tabularx}
\end{table}

\subsection{Quantitative assessment}
To objectively assess the effectiveness of the proposed multi-layer data chain system based on consortium blockchain for nuclear waste transportation tracking, a comprehensive evaluation framework must be established. Subsequently, blockchain benchmarking tools such as Blockbench and Hyperledger Caliper can be employed to conduct a thorough performance evaluation of the blockchain.\cite{fanPerformanceEvaluationBlockchain2020}

The evaluation framework should not only focus on the functional realization of the system but also quantify its performance in balancing confidentiality and transparency, a core objective. Additionally, it must comprehensively measure the system's performance, resource consumption, and security. The primary dimension of the evaluation is the effectiveness of data access control. Given the extreme sensitivity of nuclear material transportation data, the system must rigorously enforce predefined access policies. Evaluation indicators should include the rejection rate of unauthorized access attempts (target: 100\%) and the success rate of granting authorized access (target: 100\%), which can be assessed through log analysis and simulated penetration testing. Ensuring that only entities with appropriate permissions and adhering to the "need-to-know" principle can access specific data layers, particularly the operational layer, constitutes the cornerstone of system security.

The assessment framework must measure the transparency and verifiability of regulatory information, which are critical to building public trust. Key indicators include the timeliness of public layer information release (e.g., the time delay from the completion of a regulatory event to public disclosure, with target values defined by service level agreements) and the success rate of verifying public data. The latter involves simulating an audit process to confirm whether public statements, such as aggregated reports, can be reliably traced back to the records in the regulatory layer using their hash values or proof references, aiming for a verification success rate approaching 100\%.

System performance is a critical factor in assessing its practical usability. Key performance indicators include transaction throughput (TPS) and transaction latency. Throughput quantifies the system's capacity to process critical transactions, such as sensor data recording and compliance proof verification, within a given time frame. Latency measures the duration from transaction submission to its final confirmation. These metrics should be evaluated under simulated real-world network conditions and workloads using benchmarking tools like Hyperledger Caliper, and compared against expected load requirements and baseline systems. For example, the target may involve handling the aggregate rate of all sensor data during peak hours, while ensuring that latency satisfies near real-time monitoring needs.

Additionally, system overhead must be taken into account, including data storage overhead (e.g., ledger size growth rate) and computational overhead (e.g., CPU and memory usage). Monitoring these resource consumption metrics ensures the system's sustainable operation and prevents resource bottlenecks, particularly when processing high volumes of transactions or executing complex computations. Table \ref{tab:key_eval_indicators} lists the key evaluation indicators:

\begin{table}[htbp] % Placement specifier (here, top, bottom, page)
    \centering % Center the table
    \caption{Key evaluation indicators} % Table caption
    \label{tab:key_eval_indicators} % Label for cross-referencing

    % Define a new column type 'L' which is like 'X' (variable width, wrapping)
    % but left-aligned (\raggedright) instead of justified.
    \newcolumntype{L}{>{\raggedright\arraybackslash}X}

    % Use tabularx to make the table fit the text width (\textwidth).
    % All columns use 'L' type for auto-wrapping and left alignment.
    \begin{tabularx}{\textwidth}{@{} L L L L @{}} % @{} removes padding at edges
        \toprule
        % Header Row: Use \textbf for bold headers
        \textbf{Evaluation dimension} & \textbf{Index} & \textbf{Description} & \textbf{Emphasis} \\
        \midrule

        % --- Access Control Effectiveness (1 row) ---
        Access Control Effectiveness
        & Unautorized Access Attempt Rejection Rate
        & System's ability to block unauthorized access
        & Strictness of security policy enforcement \\
        \midrule % Use \midrule to separate main dimensions

        % --- Transparency & Verifiability (1 row) ---
        Transparency \& Verifiability % Use \& for ampersand
        & Public Data Verifiability Success Rate
        & Success rate of traceable verification for public information
        & Credibility of regulatory information and audit support \\
        \midrule % Use \midrule to separate main dimensions

        % --- System Performance (spans 2 rows) ---
        % \multirow{number_of_rows}{width}{content} - use '*' for natural width
        % Add \raggedright inside multirow for L-column content if needed, though L type should handle it.
        \multirow{2}{*}{\raggedright System Performance} % Spans 2 rows
        & Transaction Throughput (TPS)
        & Number of critical transactions successfully processed per unit time
        & Processing capability and efficiency \\
        \cmidrule(l){2-4} % Partial rule under sub-item, spanning columns 2-4, trimmed left (l)
        & Transaction Latency
        & Time from transaction submission to confirmation
        & Responsiveness and real-time performance \\
        \midrule % Use \midrule to separate main dimensions

        % --- System Overhead (1 row) ---
        System Overhead
        & Computational Overhead (CPU/Memory Usage Rate)
        & Computational resources required to run system components
        & Resource efficiency and scalability \\
        \midrule % Use \midrule to separate main dimensions

        % --- Security & Resilience (spans 3 rows) ---
        \multirow{3}{*}{\raggedright Security \& Resilience} % Spans 3 rows
        & Tamper Attempt Detection Rate
        & Proportion of detected tamper attempts on submitted data
        & Data integrity assurance \\
        \cmidrule(l){2-4} % Partial rule under sub-item
        & ZKP Soundness Failure Rate (if applicable)
        & Proportion of invalid ZKPs rejected by verifiers
        & Reliability of cryptographic proofs \\
        \cmidrule(l){2-4} % Partial rule under sub-item
        & System Availability (under DoS/Failure)
        & Ability to maintain service during attacks or failures
        & System robustness and reliability \\
        \bottomrule
    \end{tabularx}
\end{table}

\section{Conclusion}
This paper proposes a multi-level architecture integrating blockchain and the Internet of Things (IoT), designed to enable comprehensive tracking and management throughout the entire lifecycle of spent nuclear fuel transportation. By leveraging the immutability and distributed nature of blockchain, the automation provided by smart contracts, and the real-time data acquisition capabilities of IoT, the system effectively tackles key challenges in current spent nuclear fuel transportation management, including insufficient data transparency, stringent confidentiality requirements, and the lack of trust among collaborating parties. The hierarchical architecture in the system design successfully balances the conflicting demands of data confidentiality and regulatory transparency during transportation while fulfilling stringent security requirements. Furthermore, this paper establishes evaluation criteria for critical aspects such as data access control, transparency verification, transaction throughput, and latency. This research introduces an innovative technical framework for spent nuclear fuel transportation management, significantly improving the safety, transparency, and efficiency of the process.

%%%%%%%%%%%%%%%%%%%%%%%%%%%%%%%%%%%%%%%%%%
%\isPreprints{}{% This command is only used for ``preprints''.

%} % If the paper is ``preprints'', please uncomment this parenthesis.
%\printendnotes[custom] % Un-comment to print a list of endnotes

\bibliographystyle{unsrt}
\bibliography{Reference}

\begin{thebibliography}{10}

\bibitem{IncidentTraffickingDatabase2019}
Incident and {{Trafficking Database}} ({{ITDB}}).
\newblock https://www.iaea.org/resources/databases/itdb, April 2019.

\bibitem{huntDETERRINGNUCLEARRADIOLOGICAL}
Olympia Hunt and Stephen~V Mladineo.
\newblock {{DETERRING NUCLEAR AND RADIOLOGICAL MATERIAL THEFT}}, {{SABOTAGE}}, {{OR ILLICIT TRAFFICKING}}.

\bibitem{duRegulatoryTransparencyCitizen2023}
Juan Du and Xufeng Zhu.
\newblock Regulatory transparency and citizen support for government decisions: Evidence from nuclear power acceptance in {{China}}.
\newblock {\em Journal of Environmental Policy \& Planning}, 25(6):766--780, November 2023.

\bibitem{oecdTransparencyNuclearRegulatory2007a}
{OECD} and {Nuclear Energy Agency}.
\newblock Transparency of {{Nuclear Regulatory Activities}}: {{Workshop}} proceedings - {{Tokyo}} and {{Tokai-Mura}}, {{Japan}}, 22-24 {{May}} 2007.
\newblock Nuclear {{Regulation}}. OECD, November 2007.

\bibitem{cindyvestergaardSLAFKADemonstratingPotential2020}
{Cindy Vestergaard}, {Edward Obbard}, {Edward Yu}, {Guntur Dharma Putra}, and {Gabrielle Green}.
\newblock {{SLAFKA Demonstrating}} the {{Potential}} for {{Distributed Ledger Technology}} for {{Nuclear Safeguards Information Management}}.
\newblock Technical report, Stimson, November 2020.

\bibitem{calicSpentFuelCharacterization2022}
Du{\v s}an {\v C}ali{\v c} and Marjan Kromar.
\newblock Spent fuel characterization analysis using various nuclear data libraries.
\newblock {\em Nuclear Engineering and Technology}, 54(9):3260--3271, September 2022.

\bibitem{natarajanReprocessingSpentNuclear2017}
Rajamani Natarajan.
\newblock Reprocessing of spent nuclear fuel in {{India}}: {{Present}} challenges and future programme.
\newblock {\em Progress in Nuclear Energy}, 101:118--132, November 2017.

\bibitem{yessenbayevCombiningBlockchainIoT2024}
Olzhas Yessenbayev, Dung Chi~Duy Nguyen, Taeseok Jeong, Ki~Joon Kang, Hee~Reyoung Kim, Jonghyeon Ko, Jin-Young Park, Myung-Sub Roh, and Marco Comuzzi.
\newblock {Combining blockchain and IoT for safe and transparent nuclear waste management: A prototype implementation}.
\newblock {\em Journal of Industrial Information Integration}, 39:100596, May 2024.

\bibitem{YaoZhongJiangGuanYuQuKuaiLianYuanLiJiYingYongDeZongShu2017}
Zhongjiang Yao and JIngguo Ge.
\newblock {A Review of the Principles and Applications of Blockchain Technology}.
\newblock {\em Information Technology and Application in Scientific Research}, 8(2):3--17, 2017.

\bibitem{sharabatiBlockchainTechnologyImplementation2024}
Abdel-Aziz~Ahmad Sharabati and Elias~Radi Jreisat.
\newblock Blockchain {{Technology Implementation}} in {{Supply Chain Management}}: {{A Literature Review}}.
\newblock {\em Sustainability}, 16(7):2823, January 2024.

\bibitem{daiBlockchainBasedAccessControl2024a}
Yi~Dai, Gehao Lu, and Yijun Huang.
\newblock A {{Blockchain-Based Access Control System}} for {{Secure}} and {{Efficient Hazardous Material Supply Chains}}.
\newblock {\em Mathematics}, 12(17):2702, January 2024.

\bibitem{HarnessingPowerDistributed}
Harnessing the power of distributed ledger technology.
\newblock https://www.digicatapult.org.uk/publications/post/harnessing-the-power-of-distributed-ledger-technology/.

\bibitem{uddinSurveyAdoptionBlockchain2021}
Md~Ashraf Uddin, Andrew Stranieri, Iqbal Gondal, and Venki Balasubramanian.
\newblock A survey on the adoption of blockchain in {{IoT}}: Challenges and solutions.
\newblock {\em Blockchain: Research and Applications}, 2(2):100006, June 2021.

\bibitem{craigARGUSRFIDMONITORING2013}
Brian Craig, John Lee, Hanchung Tsai, Yung Liu, and Jim Shuler.
\newblock {{ARG-US RFID FOR MONITORING AND TRACKING NUCLEAR MATERIALS}} - {{THE OPERATING EXPERIENCE}}.
\newblock In {\em International {{Symposium}} on the {{Packaging}} and {{Transportation}} of {{Radioactive Materials}}}, San Francisco, CA, USA, 2013.

\bibitem{cosentinoSiLiFNeutronCounters2021}
Luigi Cosentino, Quentin Ducasse, Martina Giuffrida, Sergio Lo~Meo, Fabio Longhitano, Carmelo Marchetta, Antonio Massara, Alfio Pappalardo, Giuseppe Passaro, Salvatore Russo, and Paolo Finocchiaro.
\newblock {{SiLiF Neutron Counters}} to {{Monitor Nuclear Materials}} in the {{MICADO Project}}.
\newblock {\em Sensors}, 21(8):2630, April 2021.

\bibitem{ecemisExploringBlockchainNuclear2024}
Irem~Nur Ecemis, Fatih Ekinci, Koray Acici, Mehmet~Serdar Guzel, Ihsan~Tolga Medeni, and Tunc Asuroglu.
\newblock Exploring {{Blockchain}} for {{Nuclear Material Tracking}}: {{A Scoping Review}} and {{Innovative Model Proposal}}.
\newblock {\em Energies}, 17(12):3028, June 2024.

\bibitem{khutsoaneIoTDevicesApplications2017}
Oratile Khutsoane, Bassey Isong, and Adnan~M. {Abu-Mahfouz}.
\newblock {{IoT}} devices and applications based on {{LoRa}}/{{LoRaWAN}}.
\newblock In {\em {{IECON}} 2017 - 43rd {{Annual Conference}} of the {{IEEE Industrial Electronics Society}}}, pages 6107--6112, October 2017.

\bibitem{oktianHierarchicalMultiBlockchainArchitecture2020}
Yustus~Eko Oktian, Sang-Gon Lee, and Hoon~Jae Lee.
\newblock Hierarchical {{Multi-Blockchain Architecture}} for {{Scalable Internet}} of {{Things Environment}}.
\newblock {\em Electronics}, 9(6):1050, June 2020.

\bibitem{caoOverviewEdgeComputing2020}
Keyan Cao, Yefan Liu, Gongjie Meng, and Qimeng Sun.
\newblock An {{Overview}} on {{Edge Computing Research}}.
\newblock {\em IEEE Access}, 8:85714--85728, 2020.

\bibitem{InformationSecurity}
Information {{Security}}.
\newblock https://www.nrc.gov/security/info-security.html.

\bibitem{schaperLookingDemarcationNuclear2004}
Annette Schaper.
\newblock {\em Looking for a Demarcation between Nuclear Transparency and Nuclear Secrecy}.
\newblock Number no. 68 in {{PRIF}} Reports. Peace Research Institute Frankfurt, Frankfurt am Main, 2004.

\bibitem{internationalatomicenergyagencySecurityNuclearOther2024}
{INTERNATIONAL ATOMIC ENERGY AGENCY}.
\newblock {\em Security of {{Nuclear}} and {{Other Radioactive Material}} in {{Transport}}}.
\newblock {{IAEA Nuclear Security Series}}. INTERNATIONAL ATOMIC ENERGY AGENCY, July 2024.

\bibitem{chenSurveyConsortiumBlockchain2024}
Xiaotong Chen, Songlin He, Linfu Sun, Yangxin Zheng, and Chase~Q. Wu.
\newblock A {{Survey}} of {{Consortium Blockchain}} and {{Its Applications}}.
\newblock {\em Cryptography}, 8(2):12, June 2024.

\bibitem{fanPerformanceEvaluationBlockchain2020}
Caixiang Fan, Sara Ghaemi, Hamzeh Khazaei, and Petr Musilek.
\newblock Performance {{Evaluation}} of {{Blockchain Systems}}: {{A Systematic Survey}}.
\newblock {\em IEEE Access}, 8:126927--126950, 2020.

\end{thebibliography}

\end{document}